\documentclass{JHEP3}
\usepackage{amsmath,amssymb,epsfig}
\title{Cosmic censorship in Lovelock theory}

\author{Xi\'an O. Camanho\\
\sl Department of Particle Physics and IGFAE, University of Santiago de Compostela, E-15782 Santiago de Compostela, Spain\\\vskip-4mm
\email{xian.otero@rai.usc.es}}

\author{Jos\'e D. Edelstein\\
\sl Department of Particle Physics and IGFAE, University of Santiago de Compostela, E-15782 Santiago de Compostela, Spain\\\vskip-3mm
\sl Centro de Estudios Cient\'\i ficos, Valdivia, Chile\\\vskip-4mm
\email{jose.edelstein@usc.es}}

\bigskip
\bigskip
\abstract{In analyzing maximally symmetric Lovelock black holes with non-planar horizon topologies, many novel features have been observed. The existence of finite radius singularities, a mass gap in the black hole spectrum and solutions displaying multiple horizons are noteworthy examples. Naively, in all these cases, the appearance of naked singularities seems unavoidable, leading to the question of whether these theories are consistent gravity theories. We address this question and show that whenever the cosmic censorship conjecture is threaten, an instability generically shows up driving the system to a new configuration with presumably no naked singularities. Also, the same kind of instability shows up in the process of spherical black holes evaporation in these theories, suggesting a new phase for their decay. We find circumstantial evidence indicating that, contrary to many claims in the literature, the cosmic censorship hypothesis holds in Lovelock theory.}

\keywords{Lovelock gravity. Cosmic censorship. Black holes. Instabilities. Higher curvature corrections. AdS/CFT} 
\preprint{}

\begin{document}

\maketitle

\section{Introduction}

Lovelock theories \cite{Lovelock1971} are higher curvature gravities leading to second order equations of motion for the metric. The most familiar and best studied (quadratic) example is given by Lanczos-Gauss-Bonnet (LGB) gravity \cite{Lanczos}. They are higher dimensional in nature and constitute an interesting and tractable playground to explore the consequences that higher curvature terms entail for the gravitational interaction. For instance, the uses of Lovelock theory to explore several aspects of the AdS/CFT correspondence has proven to be quite fruitful \cite{Edelstein2013,Camanho2013b}.

An interesting attribute of these theories has to do with the fact that their maximally symmetric black hole geometries can be found analytically (the expressions being implicit, though) \cite{Camanho2011a} (see also \cite{Charmousis,Garraffo2008}), and their main features studied. Their dynamical response against small disturbances, captured by perturbation analysis, is of particular importance due to the strong non-linearity of the gravitational interaction. If this is already true in General Relativity, it is even more in Lovelock or other higher curvature theories. Perturbation analysis of exact solutions such as the maximally symmetric black holes alluded above, plays a crucial role in opening a window to understand the intricate dynamics of gravity in four and higher dimensions.

The application of perturbative analysis in the context of black holes was first carried out for the Schwarzschild solution by Regge and Wheeler \cite{Regge1957} in 1957, and completed a decade later by Zerilli \cite{Zerilli1970} and Teukolsky \cite{Teukolsky1972}. Stability of black holes is a fundamental issue at many levels; unstable solutions being less likely to form through a physical process and, even when they do, they will certainly not remain on that state for very long. This has many implications, from the dynamics of black holes to the determination of the final fate of gravitational collapse. Perturbation analysis also provides a criterion for uniqueness and the search for new solutions.

The stability of higher dimensional black holes in Einstein--Hilbert (EH) theory has been intensively studied (for a review see, for instance, \cite{Ishibashi2011a} and reference therein), higher dimensional Schwarzschild black holes being stable for any type of perturbations. For static spherically symmetric Lovelock black holes, the analysis is not straightforward. This is due to the complicated form of the equations of motion, even at the linearized level, but also to the form of the solutions themselves, the existence of branches, etc. Stability analyses under all type of perturbations have been performed a few years ago, though, in the case of LGB gravity  \cite{Dotti2005a,Dotti2005b,Gleiser2005,Neupane2004,Beroiz2007,Konoplya2008}. More recently, the problem was tackled in the realm of Lovelock theory by Takahashi and Soda in a series of papers
\cite{Takahashi2009h,Takahashi2009b,Takahashi2010g,Takahashi2010e}, including some connections with the AdS/CFT correspondence \cite{Takahashi2011a}.

Perturbative analysis can be formulated in the language of master equations for gauge invariant gravitational perturbations, a much simpler and intuitive approach.  In the context of Lovelock theories of gravity, generic master equations have been found in \cite{Takahashi2010g}. They cannot hide the intricacy of the theory, though. Thereby, except for some restricted efforts, most of the work has been performed in the context of LGB gravity. In this paper, we will make use of these master equations, or rather the effective potentials they define, in order to analyze the stability of black hole solutions in a particular regime, that of high momentum gravitons. This restricted analysis will greatly simplify the computations; nonetheless, it will still be general enough to uncover some very interesting features of Lovelock black hole solutions. All the computations will be performed analytically and will be useful in order to gain general intuition about gravitational instabilities in Lovelock gravities.

The most familiar instabilities occurring in Lovelock theory are of so-called Boulware-Deser (BD) type \cite{Boulware1985a}. They affect the vacua of the theory. For any Lovelock gravity, it is possible to define a formal polynomial $\Upsilon[\Lambda]$, whose coefficients are the gravitational couplings, and whose (real) roots are nothing but the plausible cosmological constant of the different vacua of the theory. A vacuum with cosmological constant $\Lambda_\star$ such that $\Upsilon'[\Lambda_\star]$ is negative, would host ghostly graviton excitations with the wrong sign of the kinetic term \cite{Camanho2013b}. If we insist in constructing a black hole solution for that vacuum, it is easy to see that the resulting configuration is unstable. Interestingly enough, there is a holographic counterpart of this result, which has to do with the unitarity of the dual CFT \cite{Camanho2010d}.

We will push the analysis of Lovelock black hole instabilities a step forward, aimed at understanding how and if the cosmic censorship conjecture \cite{Penrose} holds in these theories. There are earlier works considering its status, and there seems to be a consensus towards the idea that naked singularities can be produced via gravitational collapse in Lovelock theory. For instance, the case of LGB gravity without cosmological constant has been analyzed, both for the spherically symmetric gravitational collapse of a null dust fluid that generalizes Vaidya's solution \cite{Maeda2005} and for that of a perfect fluid dust cloud \cite{Maeda2006}. Some particular cases of Lovelock gravity have been considered as well; namely, so-called dimensionally continued gravity \cite{NozawaMaeda} and cubic Lovelock theory \cite{DehghaniF}. More recently, a broader (still greatly restricted) analysis has been tackled by several authors arriving essentially to similar conclusions \cite{Rudra,OhashiSJ2011,Zhou_etal,OhashiSJ2012}. A rough upshot of the most distinctive aspects that these papers share is:
\begin{quotation}
\noindent
{\bf (i)} The critical case --maximal higher curvature term of degree $K$ in $d_{\star} = 2K+1$ spacetime dimensions-- is qualitatively different from the $d > d_{\star}$ case.\newline\vskip-3mm
\noindent
{\bf (ii)} There is a lower bound for the mass, in the critical case, below which the spherical collapse leads to a naked singularity.\newline\vskip-3mm
\noindent
{\bf (iii)} The situation smoothens for $d > d_{\star}$, where cosmic censorship seems to be respected (to the extent it was challenged). This holds, in particular, for general relativity in higher dimensions \cite{Goswami}.
\end{quotation}
In these articles, cosmic censorship is a statement about the causal structure of spacetime, as due to the dynamical collapse. It is reasonable to question, however, whether the endpoint of the collapse, whenever it leads to a naked singularity, is stable. If it is not, the corresponding violation to the cosmic censorship hypothesis becomes dubious \cite{Joshi}. A stability analysis in the framework of gravitational collapse is cumbersome, though.

In this paper, we will tackle this problem from a different perspective. We will show that {\bf (i)} and {\bf (ii)} can be understood rather easily from a direct analysis of the spectrum of spherically symmetric static solutions, developed in \cite{Camanho2011a}. We naively focus on the possible endpoint states of the gravitational collapse by relying in Birkhoff's theorem, which is valid in this context \cite{Zegers}, and show that this leads to the same results. This allows us to extend the analysis, by the same token, to the case of planar and hyperbolic geometries. It was already seen, in the case of static planar black holes, that the Lovelock couplings have to obey certain constraints in order to avoid instabilities \cite{Camanho2010d}. For non-planar black holes the situation is more involved, but we will still be able to describe some important qualitative features of their instabilities towards uncovering their relation with the cosmic censorship hypothesis. In this way, far from being discarded as pathological, instabilities are given a very precise physical significance.

We will then discuss {\bf (iii)}. It is easy to see that the statement is wrong or, better, non-generic. It certainly applies in the cases discussed in the literature, but it is not true in a generic Lovelock theory. This is connected to the existence of two kinds of solutions, dubbed type (a) and type (b) in \cite{Camanho2011a}. While type (a) solutions are those extending all the way to the singularity, at $r=0$, as it is customarily taken for granted, type (b) solutions reach a finite radius singularity due to the existence of a critical point of the characteristic polynomial $\Upsilon$ alluded to above. This polynomial is entirely dictated by the Lovelock couplings and, as such, type (b) solutions cannot be discarded. They are pretty generic and their existence challenges the cosmic censorship hypothesis \cite{Camanho2011a}. We will nevertheless show that perturbative analysis provides valuable information on this respect, supporting its validity. We are of course aware of the many subtleties that are linked to the cosmic censorship conjecture and, in that respect, our study is not conclusive. It should be taken as a significant piece of evidence.

The paper is organized as follows. We will start by reviewing some basic facts of Lovelock theory and its spectrum of maximally symmetric static geometries. Perturbative analysis leads to a set of potentials, corresponding to the different graviton helicities. We will explore the instabilities endowed by these. We will discuss their relation to the cosmic censorship hypothesis, both for the process of black hole formation and evaporation (a curious phenomenon that might take place for spherical tybe (b) black holes, and has no parallel in the framework of general relativity), and for novel uncharged black hole solutions displaying multiple event horizons. We find circumstantial evidence indicating that, contrary to many claims in the literature, the cosmic censorship hypothesis holds in these theories. We will finally summarize and discuss our results.

\section{Lovelock black holes and graviton potentials}

Let us briefly review the black hole geometries of Lovelock theory. A fully detailed account of the bestiary of solutions has been given in \cite{Camanho2011a}. We need to recall, for the sake of completeness and fixing notation, that the action of Lovelock theory is given, in $d$ spacetime dimensions, by a sum of $K \leq [\frac{d-1}{2}]$ terms,
\begin{equation}
\mathcal{I} = \sum_{k=0}^{K} {\frac{c_k}{d-2k}} \,\mathop\int \epsilon_{a_1 \cdots a_{d}}\; R^{a_1 a_2} \wedge \cdots \wedge R^{a_{2k-1} a_{2k}} \wedge e^{a_{2k+1}} \wedge \cdots \wedge e^{a_d} ~,
\label{Ik}
\end{equation}
where $\epsilon_{a_1 \cdots a_{d}}$ is the anti-symmetric symbol, $R^{ab} := d\omega^{ab} + \omega_{\;~c}^a\wedge \omega^{cb}$ is the Riemann curvature, and $e^{a}$ is the vierbein. The gravitational couplings $c_k$ are dimensionful; we take $c_0 = 1/L^2$ and $c_1 = 1$. There is a single length scale in the microscopic theory, $L$, tantamount to the cosmological term in (\ref{Ik}). These theories admit (up to) $K$ constant curvature maximally symmetric {\it vacua}, whose {\it effective} cosmological constants are the (real) roots of
\begin{equation}
\Upsilon[\Lambda] := \sum_{k=0}^{K} c_k\, \Lambda^k = c_K \prod_{i=1}^{K} \left( \Lambda - \Lambda_i \right) ~.
\label{Upsilon}
\end{equation}
Depending upon the values of the Lovelock couplings, degeneracies may arise. We will consider a rather generic situation: If we focus on a solution whose asymptotic behavior is governed by the vacuum $\Lambda_\star$, we will demand $\Upsilon'[\Lambda_\star]$ to be strictly positive. This guarantees that the vacuum is free of BD ghosts and, furthermore, it is non-degenerate \cite{Camanho2010a}.

Among all vacua, there is one that deserves special attention: the Einstein-Hilbert (EH) branch. This is the single vacuum remaining finite in the limit in which all higher Lovelock couplings are turned off. In a vast domain of the parameter space, disconnected from the Einstein-Hilbert limiting case, it may happen that the EH-branch does not exist due to the simple fact that its would be cosmological constant is complex. This domain is referred to as the {\it excluded region} \cite{Camanho2010a}. It is fairly easy to see that the EH-branch, if it exists, does not suffer from BD ghosts.

The maximally symmetric geometries can be obtained from the following ansatz:
\begin{equation}
ds^2 = - f(r)\, dt^2 + \frac{dr^2}{f(r)} + \frac{r^2}{L^2}\, d \Sigma_{\sigma,d-2}^2 ~,
\end{equation}
where $d\Sigma_{\sigma,d-2}$ is the metric of a $(d-2)$-dimensional manifold, $\mathcal M$, of negative, zero or positive constant curvature ($\sigma =-1, 0, 1$ parameterizing the different horizon topologies). It is not difficult to verify that the Riemann tensor is singular if either the first or second derivatives of $f$ vanish. It is convenient to introduce
\begin{equation}
g(r) := \frac{\sigma-f(r)}{r^2} ~,
\end{equation}
since a general solution of the Euler-Lagrange equations can be written in terms of this function as \cite{Boulware1985a,Wheeler1986,Wheeler1986a,Aros2001}
\begin{equation}
\Upsilon[g(r)] = \frac{\kappa}{r^{d-1}} ~,
\label{LLbhsol}
\end{equation}
where $\kappa$ is proportional to the mass, $\kappa \sim M\,$V$_{\!d-2}$, with V$_{\!d-2}$ being the volume of the constant curvature manifold $\mathcal M$. This equation admits up to $K$ branches, each of them identified with a monotonic section of the polynomial. Thus, each branch ends either at $r=0$ or $r=r_{\star}$, with $\Upsilon'[g(r_{\star})] = 0$, that are precisely the points were the Riemann curvature is singular. These correspond, respectively, to the type (a)  or type (b) singularities introduced in \cite{Camanho2011a} (see Figure \ref{horrrizons}).
\FIGURE{
\centering
\includegraphics[width=0.39\textwidth]{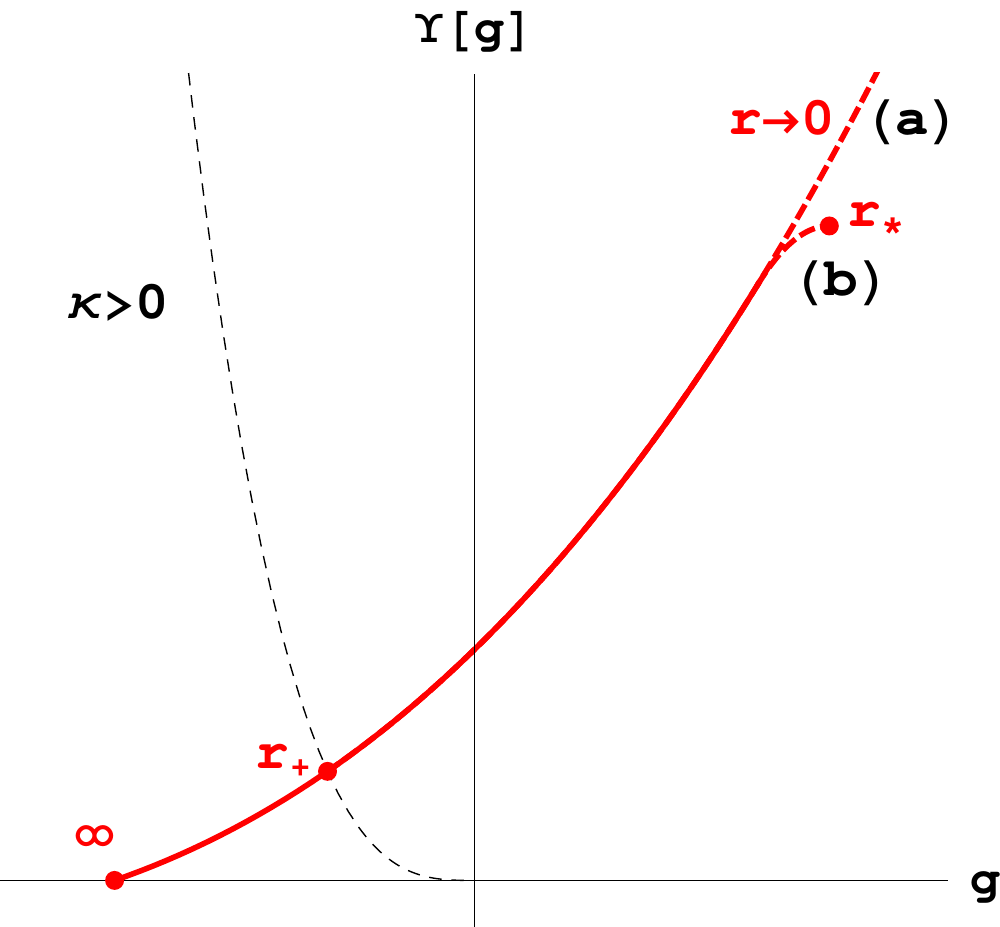} \qquad \includegraphics[width=0.39\textwidth]{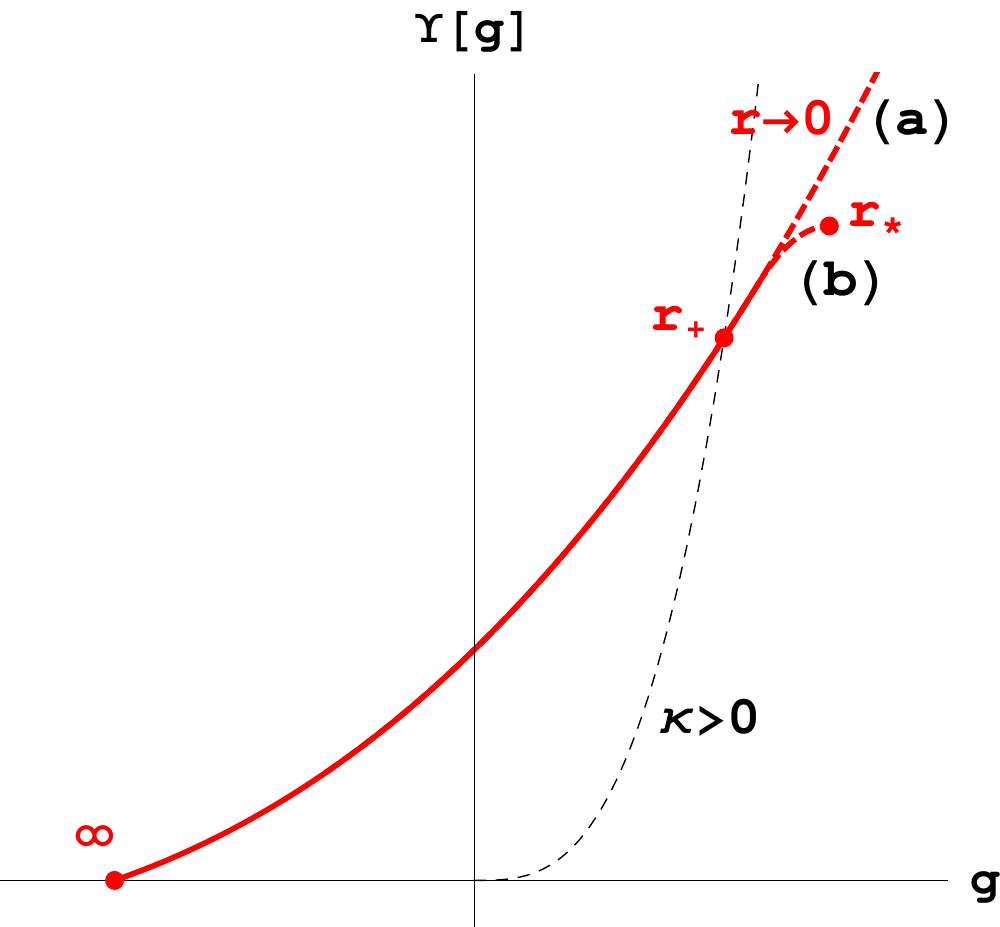}
\caption{Singularities of type (a) lie at the origin, $r=0$, while those of type (b) are finite radius and correspond to a maximum of the characteristic polynomial \cite{Camanho2011a}. The intersections with the curve (\ref{horizonnp}) give the location of the horizons for hyperbolic (left) or spherical (right) black holes.}
\label{horrrizons}}

The existence of an event horizon can be established as follows. If there is a radius, $r = r_+$, such that $f(r_+) = 0$, either $g_+ := g(r_+)$ vanishes for $\sigma = 0$, or $g_+ = \sigma/r_+^2$. The latter condition implies that $g_+$ is positive/negative for spherical/hyperbolic topology, and
\begin{equation}
\Upsilon[g_+] = \kappa\, |g_+|^{\frac{d-1}{2}} ~.
\label{horizonnp}
\end{equation}
Therefore, a given branch of the polynomial $\Upsilon$ admits an event horizon if and only if it intersects (\ref{horizonnp}). It is fairly easy to see, for instance, that the EH-branch is the only one supporting planar, as well as asymptotically AdS spherical black holes \cite{Camanho2011a}. An event horizon has an associated Hawking temperature
\begin{equation}
T = \frac{r_+}{4\pi}\,\left[(d-1)\,\frac{\Upsilon[g_+]}{\Upsilon'[g_+]}-2\,g_+ \right] ~.
\label{Temp}
\end{equation}
Notice also that Lovelock theory admits multi horizon black holes in cases where multiple intersections take place along the same branch. These solutions naively challenge some of the well established features of black holes. We shall see below that these puzzles can be resolved within the framework of this paper.

Let us consider a generic perturbation of the metric, $h_{ab}$, with fixed frequency, $\omega$, and momentum, $q$ on top of these geometries. These fluctuations split into three channels according to their polarization relative to the transverse momentum, namely the tensor, shear and sound channels \cite{Policastro2002}  (or equivalently helicity/spin two, one and zero respectively). The equations of motion for these dynamical degrees of freedom, which we call $\phi_i(r)$ for simplicity, $i=0, 1, 2$ indicating the corresponding helicity, can be recasted as Schr\"odinger type equations \cite{Takahashi2010g}, that in the large momentum limit adopt a simple form\footnote{In the notation of \cite{Takahashi2010g}, we must identify $\gamma_i \equiv \gamma = q^2 L^2$, and our potentials are related to theirs, $U_i \to - V_i/\gamma\,\Lambda$, as $\gamma \to \infty$.}
\begin{equation}
\label{Scheq}
-\hbar^2\, \partial^2_y \Psi_i + U_i(y)\,\Psi_i = \alpha^2\,\Psi_i ~,
\qquad\qquad \hbar\equiv\frac{1}{Lq} \rightarrow 0 ~,
\end{equation}
where $\alpha = \omega^2/q^2$, $y$ is a dimensionless tortoise coordinate defined as $dy/dr=\sqrt{-\Lambda_\star}/f(r)$, and $\Psi_i(y) = B_i(y)\,\phi_i(y)$, where $B_i(y)$ are functions of the metric whose specific expression can be found in \cite{Takahashi2010g}.

For a regular solution $\Psi_i(y)$, the metric perturbation $\phi_i(y)$ blows up as $B_i(y)$ approaches zero. In such case, it would not be legitimate to consider it any longer as a perturbation and, in that particular sense, the linearized analysis would be spoiled. We then need to make sure that $B_i(y)$ is non-vanishing. In our regime of interest, the effective potentials $U_i(y)$ can be determined as the speed of large momentum gravitons in constant $y$ slices,
\begin{equation}
U_i(y) = \begin{cases} {\bf c}_i^2 (y) & \qquad y < 0 ~, \cr
                      + \infty & \qquad y = 0 ~,
        \end{cases} 
\end{equation}  
$y=0$ being the boundary of the spacetime. For a generic Lovelock theory, in terms of the original radial variable, for the tensor, shear and sound channels, we find \cite{Camanho2010a}:
\begin{eqnarray}
{\bf c}_2^2(r) & = & \frac{L^2 f(r)}{(d-4)\,r^2} \frac{\mathcal{C}^{(2)}_d[g,r]}{\mathcal{C}^{(1)}_d[g,r]} ~, \nonumber\\ [0.6em]
{\bf c}_1^2(r) & = & \frac{L^2 f(r)}{(d-3)\,r^2} \frac{\mathcal{C}^{(1)}_d[g,r]}{\mathcal{C}^{(0)}_d[g,r]} ~, \label{pots}\\ [0.6em]
{\bf c}_0^2(r) & = & \frac{L^2 f(r)}{(d-2)\,r^2} \left(\frac{2\,\mathcal{C}^{(1)}_d[g,r]}{\mathcal{C}^{(0)}_d[g,r]}-\frac{\mathcal{C}^{(2)}_d[g,r]}{\mathcal{C}^{(1)}_d[g,r]}\right) ~, \nonumber
\label{GRpotentials}
\end{eqnarray}
where $\mathcal{C}^{(k)}_d[g,r]$ are functionals involving up to $k$th-order derivatives of $g(r)$,
\begin{eqnarray}
{C}^{(0)}_d[g,r] & = & \Upsilon' [g(r)]~, \\ [0.9em]
{C}^{(1)}_d[g,r] & = & \left(r \frac{d}{dr}+(d-3)\right)\Upsilon'[g(r)] ~,\\ [0.6em]
{C}^{(2)}_d[g,r] & = & \left(r \frac{d}{dr}+(d-3)\right)\left(r \frac{d}{dr}+(d-4)\right)\Upsilon'[g(r)] ~.
\label{Cpots}
\end{eqnarray}
It has been already noticed in \cite{Camanho2010a} that there is a quite simple relation between the three potentials,
\begin{equation}
(d-2)\,{\bf c}_0^2(r) - 2(d-3)\,{\bf c}_1^2(r) + (d-4)\,{\bf c}_2^2(r) = 0 ~,
\label{potsconstraint}
\end{equation}
such that any of them can be written as a combination of the other two.

These expressions are valid in general, also for charged configurations. Lovelock-Maxwell solutions were originally considered in \cite{Wiltshire1986,Wiltshire1988}. The ansatz for the metric is the same as before, whereas the field strength $F$ takes the form
\begin{equation}
F=\frac{Q}{r^{d-2}}\, e^0 \wedge e^1 ~.
\end{equation}
This solves the Maxwell equation and Bianchi identity, and sources the gravitational field in such a way that the uncharged solution (\ref{LLbhsol}) is slightly shifted to
\begin{equation}
\Upsilon[g(r)] = \frac{\kappa}{r^{d-1}} - \frac{q^2}{r^{2(d-2)}} ~,
\label{eqgCH}
\end{equation}
where $q$ is proportional to the charge $Q$, the $d$-dependent proportionality factor being positive and dimensionful. The upshot is that an implicit but exact solution, similar to the uncharged one, is found. It has an extra term in the right hand side of the polynomial equation that changes its radial dependence. Notice that the right hand side of (\ref{eqgCH}) is nothing but the $y$-intercept of $\Upsilon[g]$. Thus, once the values of the Lovelock couplings are given, which defines a concrete characteristic polynomial, the structure of the charged black hole solution and its singularities result from translating $\Upsilon[g]$ vertically but, contrary to the uncharged case discussed in detail in \cite{Camanho2011a}, this rigid translation is not monotonic, adding some degree of complexity to the problem.

The positions of the would be horizons, $r_+$, for non-planar black holes, belong to the locus given by
\begin{equation}
\Upsilon[g_+] = \kappa\, |g_+|^{\frac{d-1}{2}} - q^2\, |g_+|^{d-2} ~.
\label{rheqCH}
\end{equation}
Instead of being a monotonically increasing curve, like in the case of uncharged black holes \cite{Camanho2011a}, now $\Upsilon[g_+]$ starts growing but reaches a maximum and then decreases. The two usual horizons appearing in the Reissner-Nordstr\"om solution of the general relativistic case correspond to the pair of intersections of this curve with the characteristic polynomial. The relative slope of this intersection provides the sign of the temperature.

Let us come back to the uncharged case, which is simpler, this allowing us to make a simplifying change of variable. Instead of $r$ we take $\Upsilon$ as our independent variable (notice that the relation is one-to-one in the absence of charge), and we define $x\equiv\log L^2\Upsilon$ and $F\equiv\log \Upsilon'$, so that $r\partial_r=-(d-1)\partial_x$, yielding for the potentials\footnote{There is a common prefactor $L^2 f/r^2$ that we leave like that since it is not relevant in our discussion (it is always positive in untrapped domains of the spacetime, and vanishes at the event horizon).}
\begin{eqnarray}
{\bf c}_2^2(x) & = & \frac{(d-1)L^2 \,f}{(d-4)\,r^2}\frac{\left(\frac{d-3}{d-1}-F'(x)\right)\left(\frac{d-4}{d-1}-F'(x)\right)+F''(x)}{\left(\frac{d-3}{d-1}-F'(x)\right)} ~, \nonumber\\[0.6em]
{\bf c}_1^2(x) & = & \frac{(d-1)L^2 \,f}{(d-3)\,r^2}\left(\frac{d-3}{d-1}-F'(x)\right) ~, \label{potentialsF}\\ [0.6em]
{\bf c}_0^2(x) & = & \frac{(d-1)L^2 \,f}{(d-2)\,r^2}\frac{\left(\frac{d-3}{d-1}-F'(x)\right)\left(\frac{d-2}{d-1}-F'(x)\right)-F''(x)}{\left(\frac{d-3}{d-1}-F'(x)\right)} ~.\nonumber 
\end{eqnarray}
In this way, the potentials can be thought of as functions of the metric\footnote{Given that Lovelock theory generically possesses $K$ branches of solutions, expressions in which a given quantity depends on $r$ are hazardous due to the possibility of mixing up different solutions.} and everything can be written in terms of the Lovelock polynomial $\Upsilon$ and its derivatives; {\it e.g.}, $F'=\Upsilon[g]\Upsilon''[g]/\Upsilon'[g]^2$. As we will see, this makes easier to analyze the potentials for the different branches of solutions (corresponding to different ranges of $g$) and different values of the mass that controls the place where the solution {\it ends}, {\it i.e.}, the location of the event horizon.

\section{Black hole instabilities}

It has been found in \cite{Dotti2005b,Gleiser2005,Buchel2010a} that, in the context of the LGB theory, some effective potentials might develop negative values close to the horizon for certain values of the coupling. Therefore, given that the r\^ole of $\hbar$ in the Schr\"odinger-like equations (\ref{Scheq}) is played by $1/q$, taking sufficiently large spatial momentum we can make an infinitesimally small well to support a negative {\it energy} ($\alpha^2$) state in the effective potential. This phenomenon was also observed for Lovelock theory in higher dimensions \cite{Camanho2010d}. If we go back to the original fields, this translates into an exponentially growing and therefore unstable mode \cite{Myers2007}. Thereby, in order to analyze the stability of black holes in our regime of interest, we will be concerned just with the sign of the potentials (\ref{potentialsF}), governed by the expressions involving derivatives of $F$.

The simplest potential is the one corresponding to the shear mode. Remarkably, it has exactly the same expression as the denominators of the other two potentials. The absence of instabilities in the shear channel is then linked to the condition needed to ensure the validity of the linear analysis \cite{Takahashi2009h}, since the $B_i(r)$ functions are proportional to some power of $\Upsilon'[g]\, {\bf c}_1^2(x)$, and $\Upsilon'[g]>0$ in order to guarantee that the branch of interest is free from BD-instabilities. Thus, if we approach ${\bf c}_1^2(x) \approx 0$, the linear analysis of perturbations simply break down. There is a further reason for the shear potential to be positive. It is related to the coefficient appearing in front of the kinetic term for the gravitons and, thereby, has to be positive for the sake of unitarity. Either way, this constraint can be seen as redundant since, due to (\ref{potsconstraint}), the positiveness of the shear potential is guaranteed, as far as the tensor and sound potentials are positive. Henceforth we can restrict the stability analysis to these channels.

\subsection{Planar geometries}

We will start by considering planar geometries. This is due to their simplicity and, nonetheless, because they are the most relevant in the context of holography, their would be dual quantum field theories being defined in Minkowski spacetime. Within this framework, graviton potentials were shown to be important to probe several important aspects of the duality, ranging from causality to hydrodynamic properties of strongly coupled CFT plasmas \cite{Brigante2008a,Boer2009,Camanho2010,Boer2009a}. Planar geometries also match the high mass regime of the other two topologies, thereby providing valuable information about them. We further restrict ourselves to the EH-branch, the only one that may display a horizon in the planar case \cite{Camanho2011a}.

A negative potential well can be found at the horizon of a planar black hole geometry in a general Lovelock theory\footnote{Some work in this direction has been done recently in \cite{Takahashi2009h,Takahashi2009b} under some generic circumstances.}. Such negative values of $\mathbf{c}_i^2$ develop whenever the slope of the effective potential at the horizon is negative, and so we must require $\partial_x \mathbf{c}_i^2\leq 0$ there. Remember that $x=0$ corresponds to the horizon -- where $\mathbf{c}_i^2$ vanishes --  and $x=-\infty$ to the boundary. This is fairly simple in the planar case, where $g_+=0$, and it is straightforward to make use of $\Upsilon[g]=L^{-2}e^{x}$ to find the derivatives of $g(x)$ there, {\it e.g.},
$$
\partial_x g\big|_{g=g_+} = \frac{\Upsilon[g]}{\Upsilon'[g]} \bigg|_{g=g_+} = \frac{1}{L^2} ~.
$$
In this way, we can relate the values of the derivatives of $F(x)$ at $x=0$ with the coefficients of the polynomial $\Upsilon[g]$. The expressions for the first two derivatives appearing in  (\ref{potentialsF}) turn out to be quite simple, and are entirely given in terms of $\lambda := c_2/L^2$ and $\mu := 3 c_3/L^4$,
\begin{equation}
F'(0)=2\lambda ~, \qquad \qquad F''(0)= 2 (\mu +\lambda(1-4\lambda)) ~.
\end{equation}
Interestingly enough, higher order Lovelock couplings do not enter the relevant expressions for the stability analysis. Using these results, we expand the graviton effective potentials close to the horizon to get
\begin{eqnarray}
{\bf c}_2^2(x) &=& -x\,\frac{d-1}{d-4} \frac{\left(\frac{d-3}{d-1}-2\lambda\right)\left(\frac{d-4}{d-1}-2\lambda\right)+2(\mu +\lambda(1-4\lambda))}{\left(\frac{d-3}{d-1}-2\lambda\right)}+\mathcal O\left(x^2\right) ~, \label{horexp1} \\ [0.5em]
{\bf c}_0^2(x) &=& -x\,\frac{d-1}{d-2} \frac{\left(\frac{d-3}{d-1}-2\lambda\right)\left(\frac{d-2}{d-1}-2\lambda\right)-2(\mu +\lambda(1-4\lambda))}{\left(\frac{d-3}{d-1}-2\lambda\right)}+\mathcal O\left(x^2\right) ~. \label{horexp2}
\end{eqnarray}
The stability constraints coming from positivity of these potentials define a {\it stability wedge} in parameter space (see figure \ref{wedged}),
\begin{equation}
\mu^d_{\rm min}(\lambda) \leq \mu \leq \mu^d_{\rm max}(\lambda) ~,
\end{equation}
%
\FIGURE{
\centering
\includegraphics[width=0.85\textwidth]{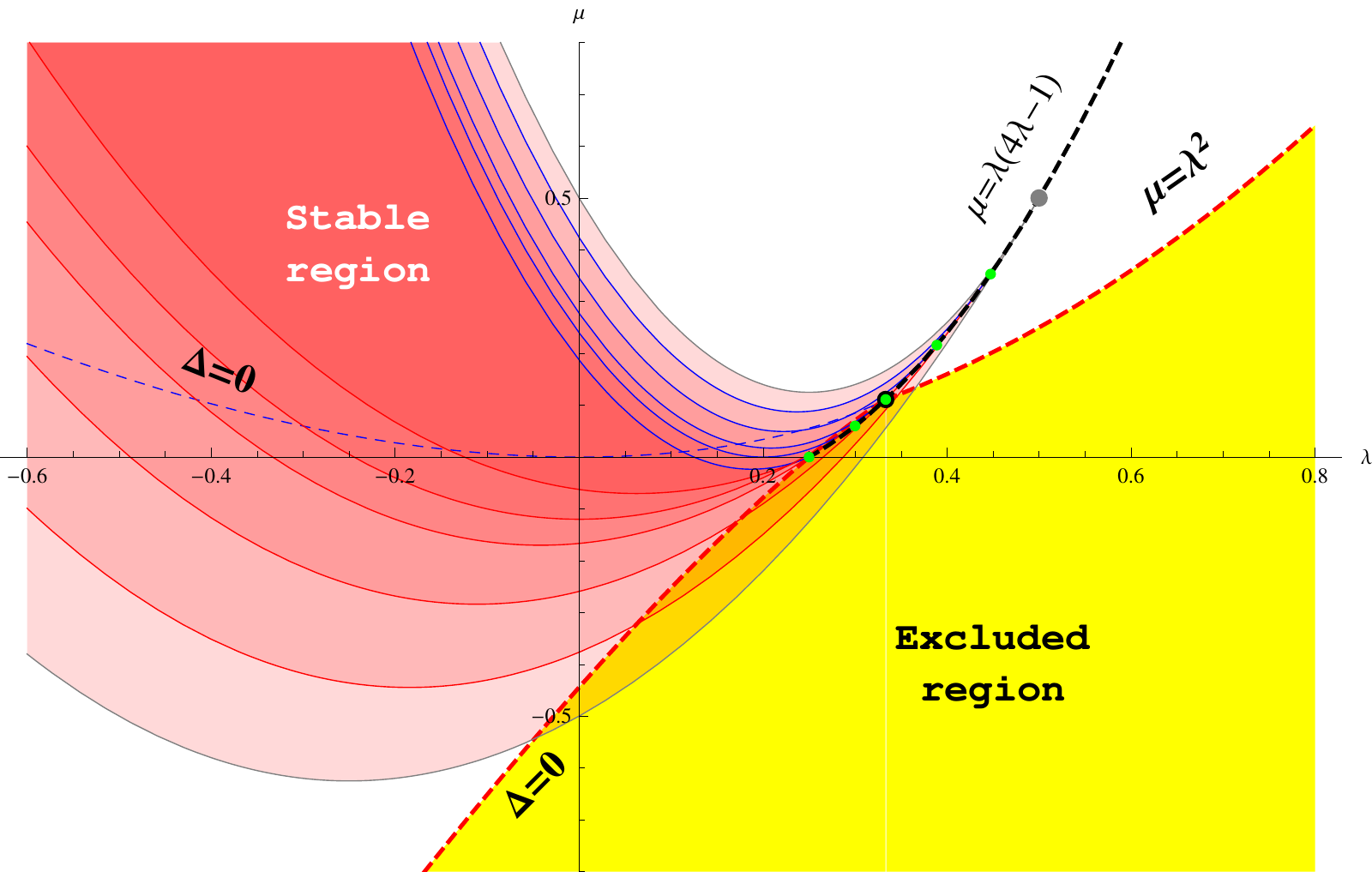}
\caption{Stability wedges (red regions) as defined by (\ref{stabwedge1}) --in red-- and (\ref{stabwedge2}) --in blue-- for $d=5,6,7,10,20$ and $\infty$ (for $d=5,6$ just the points on the $\mu=0$ axis have physical significance and bound the LGB coupling). The dimensionality increases from the innermost to the outermost curves, the gray curves that bound the faintest red region corresponding to infinite dimensionality. A given dimension stability wedge always contains those of lower dimensions. The dashed black curve describes the locus of the apex (\ref{apex}) --sequence of green dots ending at the gray dot as $d\to \infty$. Except for $d=5$ there is always a sector of the stability wedge that has to be discarded as it belongs to the {\it excluded region} (yellow area). The dashed blue and red curves are the degenerate locus $\Delta=0$ of $\Upsilon[\Lambda]$, and merge at the maximally degenerate point of cubic Lovelock theory (black dot).}
\label{wedged}}
where
\begin{eqnarray}
\mu^d_{\rm min}(\lambda) &=& - \frac{(d-3)(d-4)-2\lambda (d-1)(d-6)-4\lambda^2 (d-1)^2}{2(d-1)^2} ~, \label{stabwedge1} \\ [0.5em]
\mu^d_{\rm max}(\lambda) &=& \frac{(d-2)(d-3)-6\lambda (d-1)(d-2)+12\lambda^2 (d-1)^2}{2(d-1)^2} ~, \label{stabwedge2}
\end{eqnarray}
come, respectively, from the tensor and sound channels. For $\mu=0$, these results match those derived in \cite{Buchel2010a} for LGB gravity,
\begin{equation}
{\text d=5:} \quad  -\frac18\leq\lambda\leq\frac18 ~,
\end{equation}
the sound channel being absolutely stable in six or higher dimensions, leaving us with the constraint coming from the tensor channel,
\begin{equation}
{\text d\geq 6:} \quad  -\frac{d-6+\sqrt{5 d (d-8) + 84}}{4 (d-1)} \leq \lambda \leq \frac{-(d-6)+\sqrt{5 d (d-8) + 84}}{4 (d-1)} ~.
\end{equation}
The apex of the stability wedge is placed at the intersection of the $\lambda=\lambda_c$ line --where $\lambda_c$ is the value at which the denominator in (\ref{horexp1})-(\ref{horexp2}) vanishes-- with $\mu = \lambda (4\lambda - 1)$,
\begin{equation}
\lambda = \lambda_c = \frac{d-3}{2(d-1)} ~, \qquad \qquad \mu=\frac{(d-3)(d-5)}{2(d-1)^2} ~.
\label{apex}
\end{equation}  
In five and seven dimensions the apex coincides with the point of maximal symmetry, $\lambda=1/4$ and $(\lambda,\mu)=(1/3,1/9)$ respectively, where the polynomial $\Upsilon[g]$ has a single maximally degenerate root. For these values of the Lovelock couplings, there is symmetry enhancement and the theory becomes a {\it Chern-Simons gravity} for the AdS group (see, for example, \cite{Zanelli2005}). In higher dimensions, these values of $\lambda$ and $\mu$ define a family of theories dubbed {\em dimensionally continued gravities} \cite{BTZdim,CTZ}.

The above constraints can be interpreted under the light of the gauge/gravity correspondence as due to thermal instabilities of the dual strongly coupled fluid or plasma. In particular, the above constraints rule out negative values of the shear viscosity that would naively appear in this framework \cite{Camanho2010d}. Let us finally mention that, in addition to the event horizon stability, we may encounter negative potential wells in the bulk of the spacetime as discussed in full detail in \cite{Camanho2010d}. These are also relevant when discussing more stringent constraints on the Lovelock parameters that render the planar black holes stable.

\subsection{Non-planar geometries}

For non-planar topology the generic situation is much more involved. In addition to the gravitational couplings, the instabilities will in general depend on the parameters of the solutions, such as the mass, the charge or the radius of the black holes.\footnote{In the case of planar black holes with maximally symmetric horizons these instabilities do not depend on the mass parameter and can then be identified as a pathology of the theory itself.} However in the case of non-planar black holes these very same instabilities seem to impose constraints in the mass of the black hole. The spherical case is specially relevant as the instabilities impose a lower bound on the admissible masses for the black hole, thus they would imply that arbitrary small spherical black holes cannot form by collapse, and presumably neither the big ones as they would form gradually from smaller ones. Their values control the range of values of $g$ that correspond to the untrapped region of the spacetime where instabilities are relevant and may be found. For spherical black holes this range decreases as we increase the mass, and the opposite happens in the hyperbolic case. The planar limit is just in between, in such a way that when the planar case is unstable all spherical black holes in the EH-branch are unstable as well. Hyperbolic black holes are always stable if we decrease sufficiently the mass, and they are stable for any mass in the EH-branch when the planar limit is stable. In any case, all vacuum solutions are stable, as long as $F'=F''=0$, as well as black holes on any branch sufficiently close to them. Despite the complexity of the generic case, we may still use the stability analysis to shed light on some interesting aspects of Lovelock theory such as the evaporation of black holes and the status of the cosmic censorship hypothesis.

\section{The puzzle of multi horizon black holes}

Let us start by analyzing the case of multi horizon uncharged black holes that were mentioned earlier and are characteristic of Lovelock theory. They follow from the possibility of multiple intersections between the curve (\ref{horizonnp}) and the EH-branch in cases where the Lovelock couplings conspire to make it wiggle.
\FIGURE{
\centering
\includegraphics[width=0.43\textwidth]{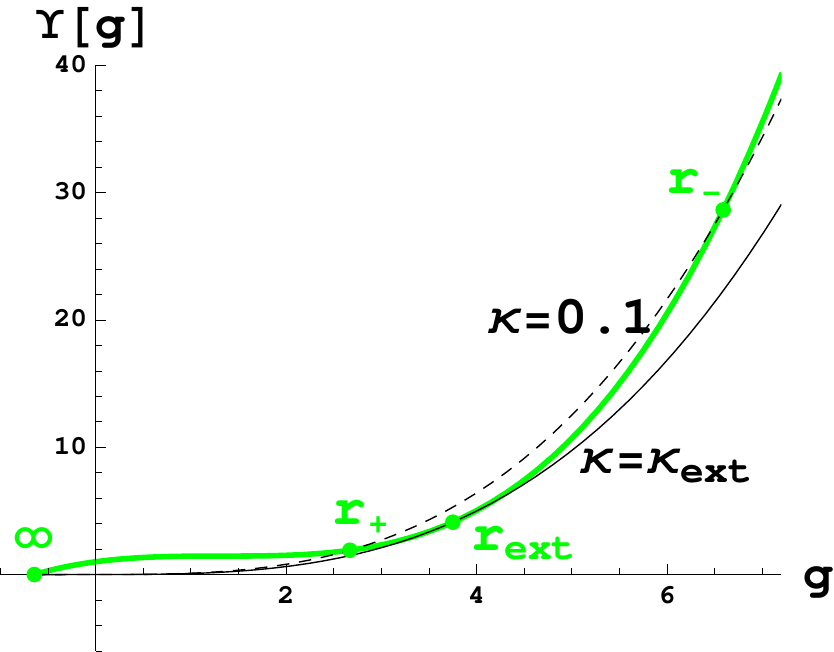}
\caption{Cubic Lovelock theory for $\lambda=-0.746$ and $\mu=0.56$. The dashed line correspond to (\ref{horizonnp}) for a spherical black hole with $\kappa = 0.1$ (in units of $L$). It crosses the polynomial twice, giving the values of $g$ at the two (outer and inner) horizons, $r_+$ and $r_-$. They would disappear for $\kappa < \kappa_{\rm ext}$ leaving a naked singularity. For $d \geq 8$, there is a third (inner) horizon, and no naked singularity, no matter how small is the mass.}
\label{extremalBHf}}
For the sake of clarity and definiteness, and without loosing generality, we will consider the cubic theory, since it contains all the ingredients necessary for the discussion. In Figure \ref{extremalBHf}, the Lovelock couplings have been chosen to display this phenomenon. The EH-branch wiggles and extends all the way to infinity. If $d \geq 8$, the solution possesses three horizons. Two are explicit in the figure, and there is a third one taking place for higher values of $g$, simply because (\ref{horizonnp}) grows at least as $g^{7/2}$, which is faster than the cubic characteristic polynomial. It is an inner horizon, but the associated temperature is positive. There are light black holes with $\kappa < \kappa_{\rm ext}$ with a unique event horizon which is precisely the alluded one. A second external horizon with vanishing temperature arises for the extremal case, $\kappa = \kappa_{\rm ext}$. Instead, heavier black holes, $\kappa > \kappa_{\rm ext}$, display this weird structure with three horizons. In the critical case, $d=7$, solutions with $\kappa < \kappa_{\rm ext}$ are naked singularities. Couples of black hole horizons appear or disappear depending upon the values of the mass parameter $\kappa$.

We can argue that the third law of thermodynamics protects the non-extremal black holes to become extremal by evaporation. It can be easily shown that the black hole would spend an infinite amount of time to reach the extremal mass. However, the real problem is the possibility of approaching these extremal solutions from lower masses. It would be possible for a non-extremal solution to accrete matter adiabatically in the right amount so that a new degenerate horizon appears. This would constitute an obvious violation of the third law. This surprising behavior also represents a puzzle in other ways. It amounts to a discontinuous change in the thermodynamic variables associated with the horizon. Moreover, we know that inner horizons are unstable \cite{Anderson1995a,Marolf2010}, so that we cannot trust our solution behind any such horizon that presumably becomes a null-spacelike singularity. Then the appearance of the new degenerate horizon would create a bigger singularity that suddenly {\it swallows} some previously untrapped region of the spacetime. Let us show how all these bizarre properties are washed out once we take into consideration the possible instabilities of these geometries.

The apparent possibility of appearance and disappearance of horizons is associated with particular points in the polynomial that solve the following constraint
\begin{equation}
\Upsilon'[g_+^{\rm ext}] = \frac{d-1}{2\,g_+^{\rm ext}}\Upsilon[g_+^{\rm ext}] ~,
\label{extremalpoint}
\end{equation}
which sets the black hole temperature (\ref{Temp}) to zero. For masses slightly above and below that point the number of horizons differs by two --a pair of consecutive outer and inner horizons merge and disappear-- and the thermodynamic variables have a discontinuity. In $d_\star = 2K+1$ dimensions, if we go below $\kappa_{\rm ext}$ the central singularity becomes naked. This however does not imply a violation of the cosmic censorship hypothesis, as long as this process is unphysical for the arguments given above.

Consider $\Xi[g] := \Upsilon[g] - \kappa_{\rm ext}\,g^{\frac{d-1}{2}}$. It is fairly easy to see that $\Xi[g_+^{\rm ext}]$ and $\Xi'[g_+^{\rm ext}]$ vanish, {\it i.e.}, $g_+^{\rm ext}$ is an {\it extremal horizon}, and $\Xi''[g_+^{\rm ext}] > 0$ (see Figure \ref{extremalBHf}), which implies\footnote{Let us further point out that there is nothing {\it a priori} preventing $\Xi''[g_+^{\rm ext}] = 0$. However, ${\bf c}_1^2(x_+^{\rm ext})$ will vanish in such case, which means that the other two potentials diverge with opposite signs, the instability being unavoidable.}
\begin{equation}
\Upsilon''[g_+^{\rm ext}] > \frac{(d-1)(d-3)}{4\,(g_+^{\rm ext})^2}\Upsilon[g_+^{\rm ext}] = \frac{d-3}{d-1} \frac{\Upsilon'[g_+^{\rm ext}]^2}{\Upsilon[g_+^{\rm ext}]} ~,
\label{extremalpoint2}
\end{equation}
if the merging horizons are an outer-inner pair.\footnote{There is another case in which the outer horizon merges with the cosmological one when we approach the Nariai mass for a dS branch \cite{Camanho2011a}. The inequality sign would be reversed in such a case.} Both conditions, (\ref{extremalpoint}) and (\ref{extremalpoint2}), make no reference to the specific value of the mass but rather express a property of the polynomial for that particular value of $g$, $g = g_+^{\rm ext}$. Taking into account that we are considering positive mass solutions, $\Upsilon[g_+^{\rm ext}]>0$, we obtain the condition
\begin{equation}
F'(x_+^{\rm ext}) = \frac{\Upsilon''[g_+^{\rm ext}] \Upsilon[g_+^{\rm ext}]}{\Upsilon'[g_+^{\rm ext}]^2} > \frac{d-3}{d-1} ~,
\label{extremalcondition}
\end{equation}
that, remarkably, is exactly equivalent to the violation of the stability condition in the shear channel,
\begin{equation}
{\bf c}_1^2(x_+^{\rm ext}) < 0 ~.
\end{equation}
As mentioned above, this negative sign also means that the graviton kinetic term on the extremal black hole background has the wrong sign, this leading to breakdown of unitarity. Notice that this phenomenon does not take place in the theory of general relativity since the second derivative of $\Upsilon$ is strictly zero in that case. We may want to interpret this result as simply telling us that the spherical extremal black hole is an unphysical solution of the Lovelock theory. Lighter black holes would also be ruled out, as long as $g_+^{\rm ext}$ belongs to their untrapped region, leading to an inconsistency. Non-extremal black holes with masses close to $\kappa_{\rm ext}$ will also be unstable under shear modes of the gravitational field.\footnote{In the more general case where several extremal points (or masses) exist, the relevant one is always the one with the biggest radius; in general, the instability would be triggered for masses slightly above it.} This interpretation would certainly solve the puzzle posed by these solutions: earlier than any jump in the black hole radius, an instability sets in driving the system somewhere else (possibly to a configuration with reduced symmetry). By the same token, this also would forbid possible violation of the third law of thermodynamics.

A final comment is in order. The previous analysis can be performed for other extremal states that appeared in the classification of Lovelock black hole solutions \cite{Camanho2011a}. These are (i) asymptotically dS spherical black holes at the Nariai mass(es) and (ii) asymptotically AdS extremal hyperbolic black holes with negative mass. They are also present in general relativity (which might seem dangerous under the light of the previous paragraph!), and it is easy to show that they respect the stability constraints at the extremal point. The value of $F'(x_+^{\rm ext})$ is not bigger but lower than the critical value of $(d-3)/(d-1)$ in these situations as either (i) $\Xi''[g_+^{\rm ext}] < 0$ (see Footnote 6), or (ii) $\Upsilon[g_+^{\rm ext}]>0$ due to the negative mass. We then do not expect instabilities for such backgrounds, though a generic stability analysis for the tensor and sound channels is not straightforward. This supports the consideration of some of these extremal solutions as groundstates \cite{Vanzo}.

\section{The cosmic censorship hypothesis and stability}

The results presented in the previous Section are relevant, as we shall see, for the cosmic censorship conjecture in Lovelock theory. For the critical dimension $d_\star = 2K+1$, at any given $K$, it may happen that the two merging horizons are the only ones and, thus, the singularity behind them would become naked. It is hard to imagine any physical process that would reduce the black hole mass below the extremal threshold without spoiling the laws of thermodynamics. Nevertheless, even if it existed, we have just shown that the instability sets in before the singularity becomes naked; in fact, before the black hole becomes extremal.

We also need to analyze the case of low mass black holes, as long as they can {\it a priori} be created directly by collapse. In order to complete the proof of their instability, we have to address the critical case, $d = d_\star$. These are naked singularities at $r=0$ (or $g \to \infty$), corresponding to type (a) branches, that take place when the highest Lovelock coupling is positive, $c_K > 0$, in the limit of small mass, $\kappa \to c_K$ (in units of $L$). But these instabilities, for $g \to \infty$, have been already observed in \cite{Takahashi2010e}, without any reference to the cosmic censorship hypothesis. In this limit, taking $F[g] := F(x[g])$,
\begin{equation}
F'[g] \approx \frac{K-1}{K} + \frac{A_K}{g^2} ~, \qquad F''[g] \approx - \frac{2 A_K}{K g^2} ~,
\end{equation}
to leading order, for some constant $A_K$ that is not relevant for our discussion. These expressions translate into the following potentials near $r = 0$:
\begin{equation}
{\bf c}_2^2(r) \approx \frac{3 L^2\,f}{(2K-3)\,r^2} ~, \qquad {\bf c}_0^2(r) \approx - \frac{3 L^2\,f}{(2K-1)\,r^2} ~.
\end{equation} 
the constraint (\ref{potsconstraint}) implying ${\bf c}_1^2(r) \approx 0$. The sound mode is unstable in accordance with \cite{Takahashi2010e}. Again, this is the case not only for the would be naked singularity but also for black holes whose mass is bigger but close enough to $c_K$. In summary, we have just shown that these solutions are unstable, a strong indication that, contrary to the claim of many papers in the literature \cite{Maeda2005,Maeda2006,NozawaMaeda,DehghaniF,Rudra,OhashiSJ2011,Zhou_etal,OhashiSJ2012}, they cannot be the end point of gravitational collapse under generic circumstances, neither be reached by any physical process.

For matter collapsing into a regular black hole with an event horizon, the formation of the latter constitutes a critical moment for the matter contained within it. Think of a spherically symmetric configuration. The causal properties of event horizons force all the matter to end up at the central singularity, regardless of the details of its distribution, and it cannot escape, since this would be nothing but traveling backwards in time. This is radically different if the would be end point is a naked singularity. Not being dressed with an event horizon, nothing prevents matter from escaping the singularity without violating any causal structure. Thereby, the final configuration may be much more sensitive to the details of the mass distribution under collapse. The Penrose diagrams corresponding to both such processes are schematically depicted in Figure \ref{collPenrose}.
\FIGURE{
\centering
\includegraphics[width=0.73\textwidth]{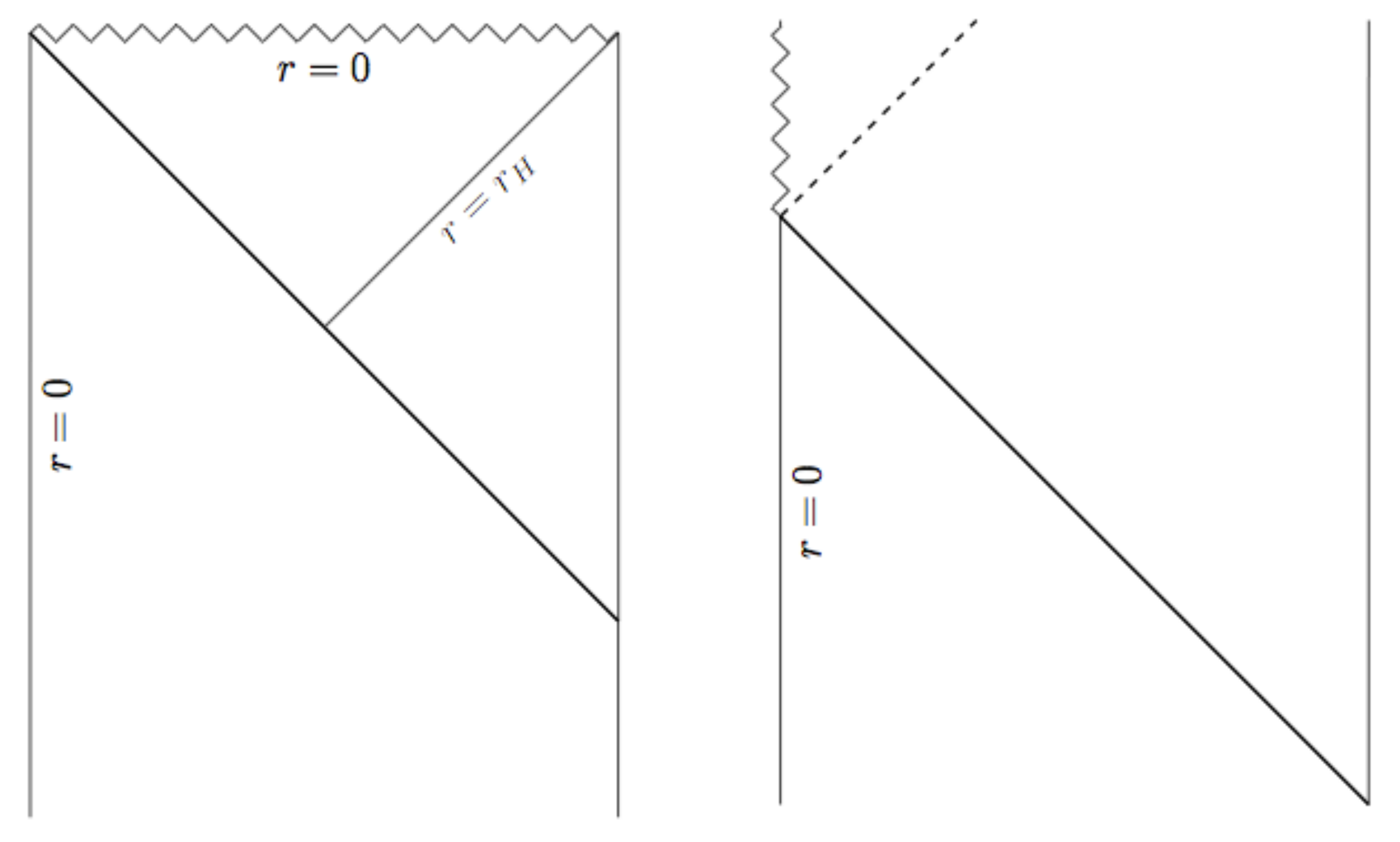}
\caption{Penrose diagrams for the collapse of a shell of radiation (thick line) to a black hole (left) and a naked singularity (right). In the case of the naked singularity the radiation has no obstacle to escape {\it across} (or bouncing on) the singularity, the hypothetical trajectory corresponding to the dashed line (that is also a Cauchy horizon).}
\label{collPenrose}}
It seems reasonable to use the stability properties of such hypothetical solutions in order to assess whether or not they represent good candidates for endpoints of gravitational collapse.

We have just shown that such a solution is unstable and, most probably, any small departure from spherical symmetry would imply that such a naked singularity is not going to be formed. Even in the spherically symmetric case, the singularity might just be {\it spurious}, mathematically a result of all the matter ending up at the same point at the same time due to the rigidity of such a symmetric ansatz. Furthermore, once it reaches the singularity, matter may still bounce back even in presence of other non-gravitational interactions. This is clear if we distort slightly the matter configuration in such a way that we avoid this {\it coincidence problem}. Different parts of the matter configuration will arrive at different times at slightly different points in such a way that the singularity is not formed; {\it e.g.}, if we provide some angular momentum, the centrifugal barrier would do the job, at least in some cases.
\FIGURE{
\centering
\includegraphics[width=0.47\textwidth]{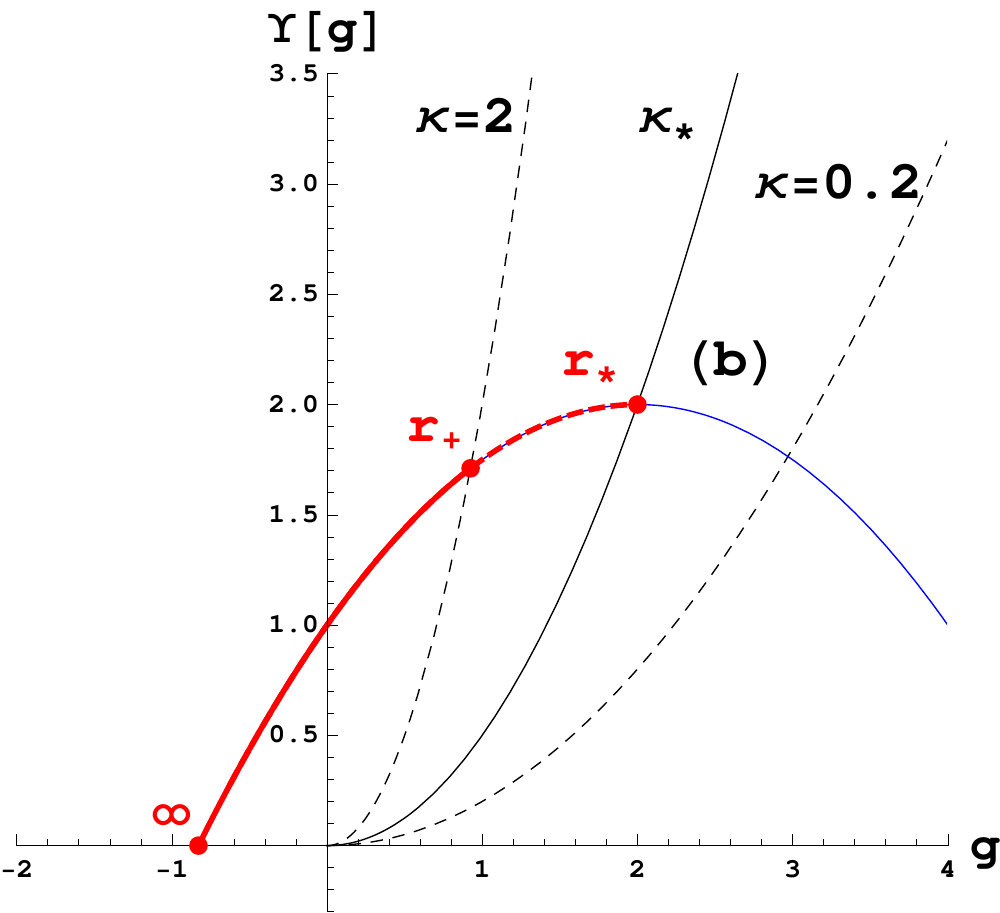} 
\caption{Spherical black holes in the EH-branch of the type (b). The singularity is naked for $\kappa \leq \kappa_{\star} = 0.5$ (in $L$ units). In the limiting case, the event horizon shrinks to the singularity point $r_\star$, $g_+ = g_\star := g(r_\star)$.}
\label{relevantGB}}

There is however a substantially different kind of naked singularity that belongs to the realm of Lovelock theory (possibly to other higher curvature gravities as well). It is quite generic and can be formed through seemingly plausible physical processes. It may occur for the otherwise well-behaved EH-branch, that is the main focus of the present paper, as well as for the {\it higher curvature} AdS/dS branches. They arise due to the fact that the branch of interest ends up at a maximum of the Lovelock characteristic polynomial, a finite radius, $r_\star = r^{-1}(g_\star)$, with $\Upsilon'[g_\star] = 0$, before an event horizon is encountered,\footnote{Another type of singularity is that associated with positive minima of $\Upsilon$. They would correspond to a branch with a complex cosmological constant. This singularity can never be avoided by any choice of mass parameter and is part of what we have called, in the case of the EH-branch, the excluded region.} a so-called type (b) branch (also AdS branches fit into this behavior) in the classification of \cite{Camanho2011a}; see Figure \ref{relevantGB}.

The above discussions make clear that the stability of these solutions is an essential point to be analyzed in order to unravel the status of the cosmic censorship hypothesis in these theories. In fact, we will see that these instabilities show up whenever a violation of the cosmic censor may take place. It is enough to probe the solution close to the singularity in order to show its instability. We just need to analyze the behavior of the potentials for values of $g$ close to the critical one, $g_\star$. Given that it is a local maximum of $\Upsilon$, we can approximate the polynomial by
\begin{equation}
\Upsilon[g] \approx \Upsilon[g_\star] + \frac12 \Upsilon''[g_\star]\, (g-g_\star)^2 ~,
\end{equation}
and use this to compute the leading contribution to the derivatives of $F[g] := F(x[g])$, that diverge in the vicinity of the singularity,
\begin{equation}
F'[g] = \frac{\Upsilon[g] \Upsilon''[g]}{\Upsilon'[g]^2} \approx \frac{\Upsilon[g_\star]}{\Upsilon''[g_\star]\,(g-g_\star)^2} \to - \infty ~, \qquad F''[g] \approx -2 F'[g]^2 ~,
\end{equation}
the negative sign in $-\infty$ coming from the fact that $g_\star$ is a maximum of $\Upsilon$. Consequently, the tensor and sound potentials diverge with opposite sign,
\begin{equation}
{\bf c}_2^2 \approx \frac{(d-1)L^2\,f_\star}{(d-4)\, r^2}F'~, \qquad {\bf c}_0^2 \approx \frac{(d-1)L^2\,f_\star}{(d-2)\, r^2}(-3F') ~.
\end{equation}
this being a clear sign of unavoidable instability. Again, since we have expanded around $r_\star$, not only the naked singularity is unstable but any black hole solution whose horizon is close enough to it, $r_+ \simeq r_\star$; {\it i.e.}, $\kappa \gtrsim \kappa_\star$. Hence, the black hole solutions cannot be continuously connected to the singular solution by any physical process since the instability would inevitably show up with increasing importance before the singularity is attained, driving the system somewhere else. This is an important point, as long as the threshold between type (b) black holes and naked singularities is not an extremal solution.

\section{Instabilities and black hole evaporation}

In our quest to qualitatively understand in which situations Lovelock black holes become unstable, we have to review the case of evaporating black holes. It is relevant as in previous analysis the instabilities seemed to appear when the black hole horizon approaches {\it too much} the singularity. This is for instance what happens for type (b) spherical solutions. As the black hole evaporates it looses mass approaching the critical value (for which the temperature diverges) in finite time. We have also seen that for type (a) branches in $d=2K+1$ the black hole solution becomes unstable before reaching the $r_+=0$ state, that in this case is extremal. This process would leave behind a naked singularity, but we already proved that neither the extremal solution nor the naked singularity can be reached, the solution becoming unstable as the horizon gets {\it close} to the singularity. The black hole would spend an infinite amount of time to become extremal but just a finite lapse to reach the instability. Hence, the instability seem to play a r\^ole in the evaporation process of black holes in Lovelock gravities.

The remaining case to be understood is that of type (a) solutions, either belonging to the EH or dS branches, for dimensions bigger than the critical one, $d>2K+1$. As the mass of these black hole shrinks to zero, the horizon also contracts approaching the central singularity. The singularity can never become naked in this case but, still, instabilities show up before the black hole shrinks to zero size. In \cite{Takahashi2009b,Takahashi2010e}, Takahashi and Soda observed that when all possible Lovelock couplings are turned on --the highest one, $c_K$, being positive--, the instability always shows up, in even and odd dimensions. Here we will generalize that analysis.

Expanding the tensor and scalar potentials around $g\rightarrow\infty$ we simply get
\begin{equation}
{\bf c}_2^2 \approx \frac{(d-3K-1)L^2\,f}{K(d-4)\,r^2} ~, \qquad {\bf c}_0^2 \approx \frac{(d-K-1)L^2\,f}{K(d-2)\,r^2} ~.
\end{equation} 
Therefore, any spherical branch under the above conditions will present a tensor instability in the small mass limit for $2K+1 < d < 3K+1$, the sound channel being always well behaved. The $d=3K+1$ is special and we need to go to the following order in the tensor potential, and whether the black hole is stable or not will depend on the actual values of $(c_{K-1}/c_K)^2$ and $c_{K-2}/c_K$. The Einstein-Hilbert theory is stable in any number of dimensions.

Summarizing, for generic Lovelock theory we have encountered instabilities at the end of the evaporation process of any type (a) spherical black hole in dimensions lower than $3K+1$, whereas for type (b) the instability appears for any spacetime dimensionality, regardless of the asymptotics. The latter dimensionality, $d=3K+1$, has to be considered more carefully. Whether this instability is pointing towards an inconsistency of the theory that forces us to constrain $K \leq\left[\frac{d-1}{3}\right]$ or, else, it can be ascribed to a physically sensible process taking place during the black hole evaporation has yet to be answered.

\section{Final remarks and discussion}

Perturbative analysis of exact solutions appears to be an extremely useful tool to get a deeper understanding of the dynamics of gravity in four and higher dimensions. We have presented and analyzed the graviton potentials for generic Lovelock gravities in a particularly simple regime, that of very large momentum and frequency. Still, despite the simplicity of the adopted approach, the results are general enough to provide some valuable information about the behavior of black holes in these higher curvature theories.

Lovelock solutions were shown to display several seemingly pathological features, ranging from naked singularities to violations of the third law of thermodynamics or the existence of discontinuous changes on the radius of multi horizon black holes and, consequently, also on their associated thermodynamic variables. We have shown that all the puzzling properties of these solutions are ruled out once their stability is considered. Before any naked singularity may show up, the corresponding solution becomes unstable, this applying, to the best of our understanding, both to black holes undergoing any physical process --accretion of matter or evaporation--, as well as to the collapse of any kind of matter. Therefore, naked singularities cannot be formed under the evolution of generic initial conditions. The existence of such solutions would require extremely {\it fine-tuned} initial data for the gravitational collapse. Any perturbed set of initial conditions will not end up with the formation of a naked singularity due to the instability. To the extent we could challenge it, the cosmic censorship hypothesis seems at work in Lovelock theory. This seems relevant from the point of view of the uses of Lovelock gravities in the framework of the AdS/CFT correspondence.

These results provide a new insight into these graviton instabilities, too often taken as pathological. Much on the contrary, they seem to acquire physical significance and play an important r\^ole, healing Lovelock theory from what otherwise would lead to an ill-defined behavior. Their analysis is crucial for the understanding of the dynamics of black holes in Lovelock theory and possibly other theories containing higher powers of the curvature. In particular, we have shown that Lovelock black holes with spherical symmetry generically become unstable as they evaporate. In most cases there is a mass gap between the lightest stable black hole and the corresponding maximally symmetric vacuum. The instability may be a hint pointing to the existence of new solutions that fill this gap. The only possibilities we foresee for such hypothetical states are {\it stars} made either of regular matter or {\it gravitational hair}.\footnote{See \cite{Dias2011} for an example of the class of solutions we are referring to.} In some cases we might need to break spherical symmetry and provide some angular momentum, as the perturbations considered do.

We have focused mostly on spherical symmetry and the EH-branch in this paper. The results can be straightforwardly extended to AdS/dS-branches. In the former case, hyperbolic symmetry is the relevant one, and many results mirror those of spherical black holes in dS-branches. For hyperbolic AdS-branches the stability of the black hole solutions provides an upper bound on the gravitational mass coming, again, from maxima of the characteristic polynomial. The bound should not be of fundamental nature, thus we expect more massive solutions in the form of hairy black holes, the {\it dressing} of the horizon providing the extra required energy. The fact that the instability is restricted to a finite (radius) spacetime region seems to support this hypothesis. This is the region filled by our matter configuration or, better, our lump of gravitational energy or {\it geon}, the rest of the spacetime remaining untouched. A more detailed analysis is necessary in order to confirm or disprove this intuition.

Coming back to the spherical case,\footnote{The hyperbolic case can be discussed quite straightforwardly by mirroring to negative values of $g$. There are some subtleties, though, that are beyond the scope of this paper \cite{CamanhoPhD}.} the instability is associated with a particular value of $g$, $g_\ast < g_\star$, above which some of the potentials become negative. The threshold of the instability is then the mass $\kappa_\ast$ for which the radius of the horizon is $r_\ast = g_\ast^{-1/2}$. For lower values of the mass, we can translate the stability constraints into a bound on the amount of matter that can be contained in a hypersphere of radius $r$. For any quantity of matter contained in such a hypersphere, with mass $\kappa(r)$, the radius should be bigger than the would be unstable region for the same mass. For a continuous distribution of matter, this has to be verified for all values of the radius down to $r=0$ so that the configuration is stable. The equation for the metric function $g$ should be schematically modified \cite{Paulos2011} in that case to
\begin{equation}
\Upsilon[g(r)] = \frac{\kappa(r)}{r^{d-1}} \leq \Upsilon[g_\ast] ~.
\label{denssity}
\end{equation}
Of course, to fully address this issue we would need to consider adding matter fields that in principle may change the stability analysis. The characteristic polynomial plays the r\^ole of an {\it effective density} and its value at the threshold of the instability can be interpreted as the maximal one so that the instability is avoided; see Figure \ref{stars}.
\FIGURE{
\centering
\includegraphics[width=0.67\textwidth]{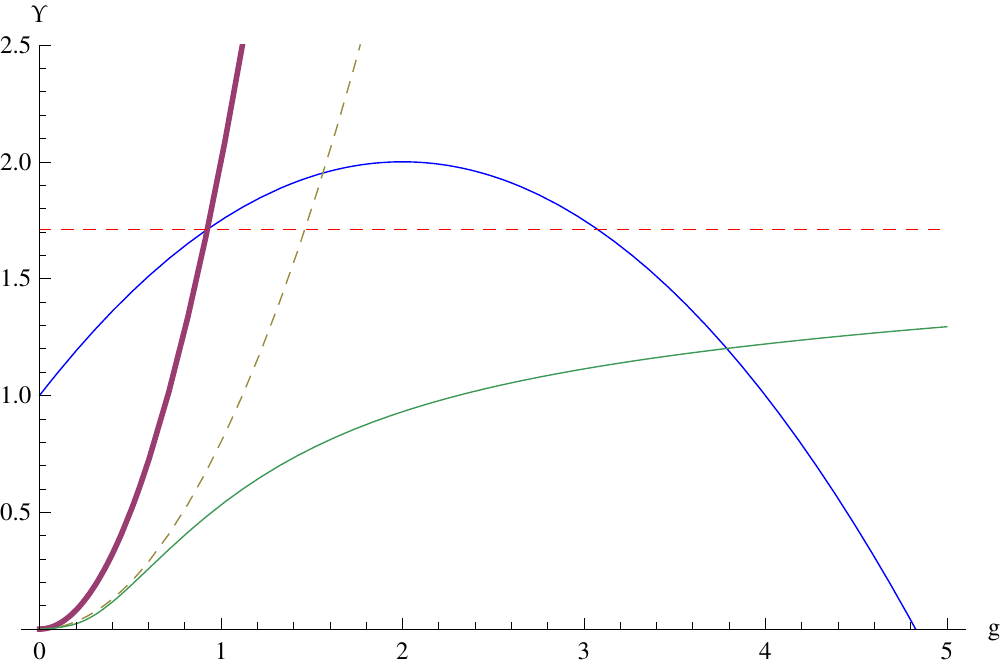}
\caption{Possible resolution of the instabilities in the spherical case for the EH-branch. For masses below $\kappa_\ast$ (violet curve), the polynomial is interpreted as an effective density with $\kappa(r)$ being such that (\ref{denssity}) holds. This {\it star}-like solution is represented by the green curve; it has no event horizon but smoothly reaches the origin. The type (a) case admits an analogous explanation, since the behavior of $\Upsilon$ for $g > g_\ast$ is irrelevant.}
\label{stars}}

Even in the event that these configurations are not possible in a Lovelock theory coupled to matter, there is another possible endpoint for gravitational collapse. The collapsing matter may just disperse again as it is not trapped by any event horizon. In case the symmetry is relaxed, the matter will not collapse exactly to one point in such a way that the density may remain finite through the evolution. Then it may disperse back to infinity or form some type of bound state. To be conclusive in this point it seems indispensable to explore the inclusion of matter.

It would be interesting to investigate the status in Lovelock theory of the nonlinear instability of AdS that was found in General Relativity under small generic perturbations \cite{Bizon2011}. Such an instability is triggered by a mode mixing that produces energy diffusion from low to high frequencies. This mechanism can be understood as a result of the interaction of geons that end up producing a small black hole \cite{Dias2011}. Given the existence of a mass gap in the Lovelock black hole spectrum, it is natural to wonder whether AdS does not display turbulent instabilities when small generic perturbations are considered within the higher curvature dynamics. The only piece of information on this direction that we are aware of is the analysis of Choptuik's critical phenomenon \cite{Choptuik} in the case of LGB theory without cosmological constant \cite{Golod}.

A final comment is in order. There is a different notion of instability that may supersede that unraveled by perturbative analysis in this paper. Black hole solutions are subjected to the laws of thermodynamics and, as such, they may become unstable at the threshold of a (for instance, Hawking-Page) phase transition. In a thermal bath, it has been recently shown, indeed, that Lovelock theory admits generalized phase transitions allowing for jumps between the different branches \cite{Camanho2012,Camanho2013}. This phenomenon further enriches the rules of the game and make it cumbersome and interesting to reinstate a basic question with which we would like to finish the paper: Are higher curvature gravities like Lovelock theory fully consistent?

\section*{Acknowledgements}

We wish to thank \'Oscar Dias, Ricardo Monteiro and Jorge Santos for useful discussions.
This work was supported in part by MICINN and FEDER (grant FPA2011-22594), by Xunta de Galicia (Conseller\'{\i}a de Educaci\'on and grant PGIDIT10PXIB206075PR), and by the Spanish Consolider-Ingenio 2010 Programme CPAN (CSD2007-00042).
We would like to thank the Universities of Buenos Aires and Andr\'es Bello, as well as the Institute for Advanced Study and the Benasque Centre for Science Pedro Pascual for hospitality during part of this project.
X.O.C. is thankful to the Front of Galician-speaking Scientists for encouragement.
The Centro de Estudios Cient\'{\i}ficos (CECs) is funded by the Chilean Government through the Centers of Excellence Base Financing Program of Conicyt.


\end{document}